\documentclass[12pt]{iopart}
\pdfoutput=1
\usepackage{graphicx}
\usepackage[latin1]{inputenc}
\usepackage{amsfonts}
\usepackage{amssymb}
\graphicspath{{./images/}}
\begin{document}
\title[Metabolic turnover from EBA]{Genome-scale estimate of the metabolic turnover of E. Coli from the energy balance analysis.}
\author{D.De Martino$^{1}$}
\address{$^1$ Center for life nanoscience, Istituto Italiano di Tecnologia, CLNS-IIT,  Viale Regina Elena 291, 00161, Rome, Italy}
\begin{abstract}
In this article the notion of metabolic turnover is revisited in the light of recent results of out-of-equilibrium thermodynamics. By means of Monte Carlo methods we perform an exact uniform sampling of the steady state fluxes in a genome scale metabolic network of E Coli from which  we infer the metabolites turnover times.
However the latter are inferred from  {\em net} fluxes, and we argue that this approximation is not valid for enzymes working nearby thermodynamic equilibrium.
We recalculate turnover times from {\em total} fluxes  by performing an energy balance analysis of the network and recurring to the fluctuation theorem. We find in many cases values one of order of magnitude lower, implying a faster  picture of intermediate metabolism. 
\end{abstract}

\maketitle
\section{Introduction}
The recent wealth of data coming from genome sequencing in  biology\cite{marx2013biology} is eager for unifying schemes, interpretations and insights that could come from physics.  
In particular metabolism, the ubiquitous and highly conserved enzymatic network devoted to free energy transduction in every cell, has been the subject of structural reconstructions at the scale of the whole genome\cite{palsson2015systems}. 
Metabolism has deep physical roots and thus a long standing tradition of physical modelization efforts\cite{heinrichregulation}.
A current challenge faced by physics thus concerns  the extension of such efforts in large scale models.
On one hand we lack detailed information on many parameters, 
on the other, even simple models with minimal assumptions lead to difficult computational issues. Much attention has been payed to the structural properties of metabolic networks
\cite{braakman2013compositional}, but   
on the other hand metabolism is inherently dynamical  
and a fundamental inherently physical question 
regards the assessment of its typical timescales.
In this article we will analyze metabolites turnover times at  the scale of the whole genome in a metabolic network model of E Coli. 
We will point out that such analysis requires to integrate thermodynamic information and thus the evaluation of an energy balance analysis of the network\cite{Beard:2002vn, babson}.\\ 
If we model a metabolic  system in terms of the dynamics of the concentration levels, assume well-mixing (no space) and neglect noise (continuum limit), we still have a very large non-linear dynamical system whose parameters can be unknown.  
For a chemical reaction  network  in which $M$ metabolites participate in $N$ reactions  with the  stoichiometry encoded in a matrix $\mathbf{S}=\{S_{\mu r}\}$, 
the concentrations $c_\mu$ change in time according to mass-balance equations
\begin{equation}
\dot{\mathbf{c}} = \mathbf{S \cdot v}
\end{equation}
where $v_i$ is the flux of the reaction $i$ that is in turn a  function of the concentration levels  $v_i(\mathbf{c})$.
Upon considering a steady state (homeostasis), i.e. a flux configuration satisfying 
\begin{equation}
\mathbf{S \cdot v}=0
\end{equation}
we could in principle determine rigorously the timescales by performing a linear stability analysis of these steady states, i.e. upon linearizing the laws $v_i(\mathbf{c})$ and finding  the spectrum of the resulting matrix.
 Such calculation requires knowledge of the elasticity coefficients\cite{fell1997understanding} $\frac{\partial v_i}{\partial c_\mu}$ and in turn of the reaction laws with their parameters, that is not the case in large scale models.
A widely employed approximation, if at least fluxes and concentrations are known, is to consider the metabolites turnover times\cite{easterby1973coupled,easterby1981generalized} $\tau$, i.e. the ratio between the concentration of a given metabolite $c$ and the flux of production $P$ (or equivalently destruction $D$, that is the same in the steady state), i.e.
\begin{equation}
\dot{c} =P -D = 0, \quad \tau =\frac{c}{P} 
\end{equation}
This turnover time is usually  directly and intuitively interpreted as the average time it takes to fully replenish a
given metabolic pool.
However, we point out that, especially nearby thermodynamic equilibrium, net fluxes result from the difference between forward and backward contributions
$\nu=\nu^+-\nu^-$. This would imply that production and destruction fluxes split as well and the resulting  turnover time could be lower\cite{reich1981energy}:
\begin{eqnarray}
\dot{c} =(P^+ + D^-) -(D^+ + P^-) = 0 \nonumber \\
\tau =\frac{c}{P^+ + D^-} 
\end{eqnarray}
It should be understood that these contributions directly affect even the calculation of relaxation times from a more rigorous linear stability analysis. 
Now, if we have  information about the  net flux $v$ and the free energy $\Delta G$, we can estimate the backward and forward contribution $v^+, v^-$ from a simplified form\cite{beard2007relationship} of one of the main result in out-of-equilibrium thermodynamics, the fluctuation theorem\cite{gallavotti1995dynamical, gaspard2004fluctuation,schmiedl2007stochastic}:
\begin{equation}
\frac{v^+}{v^-} = e^{-\Delta G/RT}
\end{equation} 
It has been shown that this simplified form is valid for the mass action law and the reversible Michellis-Menten kinetics, and it has been proposed to be valid in general\cite{beard2007relationship}.
For instance, consider a metabolite in the red cell, Glucose-6-phosphate, produced in glycolysis\cite{berg2002glycolysis} by an irreversible reaction, Hexokinase ($\Delta G_1 \simeq -29$ KJ/mol),  and consumed by a reversible one, phosphoglucoisomerase ($\Delta G_2 \simeq -2.9$ KJ/mol). We can calculate at $ RT = 2.5$KJ/mol that the  turnover time $\tau_0$ from net fluxes overestimates the one $\tau$ from total fluxes by a factor
\begin{equation}
\frac{\tau_0-\tau}{\tau} \simeq \frac{1}{e^{-\frac{\Delta G_2}{RT}}-1}\simeq 45\%
\end{equation}
In the following we will show the results of an uniform sampling of the possible steady state fluxes in a genome scale metabolic network model for the bacterium E Coli.  Then, the results of an energy balance analysis of the network will be presented in order to  estimate 
total fluxes from the simplified form of the aforementioned fluctuation theorem. 
We will calculate metabolites turnover times from net fluxes,  correct them from estimate of total fluxes and show that in the latter case they can be much lower.  We will finally draw out some conclusions, for instance regarding how this new faster picture of the intermediate metabolism affects the well-mixing hypothesis.

\section{Results and discussion}

We consider the steady state fluxes of the metabolic network model of E Coli iJR904\cite{reed2003expanded} in a glucose-limited minimal medium (see materials and methods sec.).

In constraint based modeling, apart from mass balance constraints, fluxes are bounded in certain ranges $v_r \in [v_{r}^{{\rm min}},v_{r}^{{\rm max}}]$ that take into account thermodynamical irreversibility, kinetic limits and physiological constraints.
The set of constraints
\begin{eqnarray}\label{eq3}
\mathbf{S \cdot v}=0, \nonumber \\
v_r \in [v_{r}^{{\rm min}},v_{r}^{{\rm max}}]
\end{eqnarray}  
defines  a convex closed set in the space of reaction fluxes: a polytope from which feasible steady states can be efficiently inferred with Monte Carlo methods\cite{Wiback:2004kc} (see materials and methods sec.)

Once we have the flux distributions, we can single out for each metabolite the net production flux, that is the sum of positive definite terms (and that it is equal to the net consumption flux under our steady state assumption), if we have information about the concentration levels we can thus calculate the turnover times, i.e.
\begin{eqnarray}
\dot{c}_\mu &=& P_\mu -D_\mu = 0 \nonumber \\
P_\mu &=& \sum_{i}\theta(S_{i \mu} v_i)S_{i \mu} v_i \nonumber \\
D_\mu &=& -\sum_{i}\theta(-S_{i \mu} v_i)S_{i \mu} v_i \nonumber \\ 
\tau_\mu &=&\frac{c_\mu}{P_{\mu}} 
\end{eqnarray}

\begin{figure}[h!!!!!]\label{fig1}
\begin{center}
\includegraphics*[width=0.7\textwidth,angle=0]{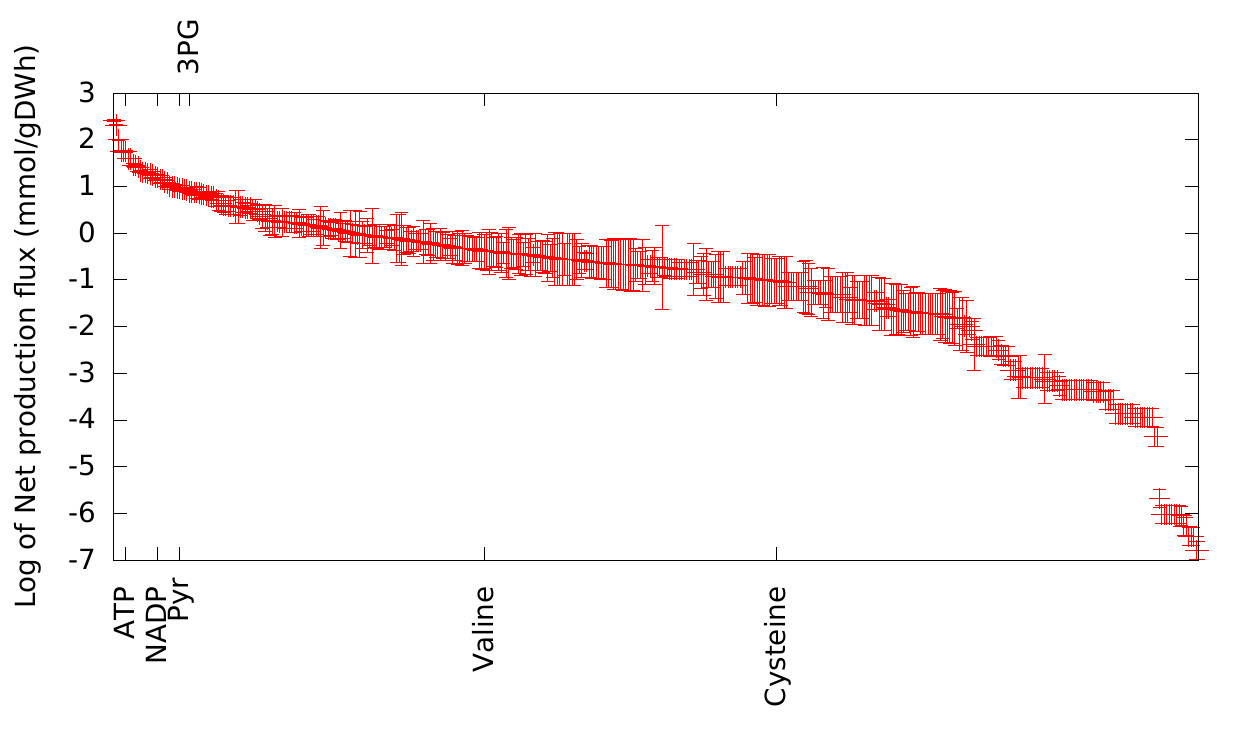}
\includegraphics*[width=0.7\textwidth,angle=0]{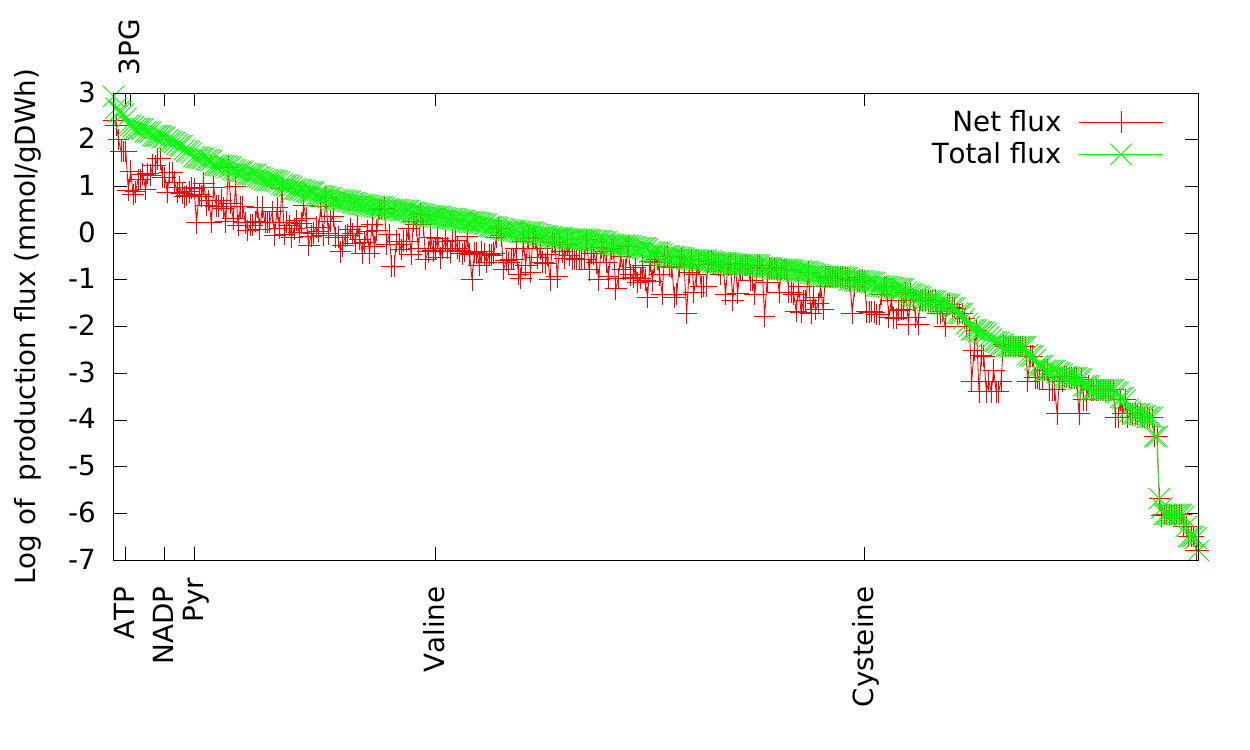}
\caption{Top: Net metabolites production fluxes from uniform sampling of steady states of the metabolic network model of E Coli iJR904 in a minimal medium. Bottom: Total fluxes from energy balance analysis, compared with net fluxes.}
\end{center}
\end{figure}
We show in fig 1 (top) in ordered fashion the mean of such production fluxes in log scale (for the sake of avoiding confusion we indicate the name of only some metabolites on the x axis). They span  $8$ orders of magnitude ranging from   $10^{1-2}$ to $10^{-7}$
mmol/gDWh, thus giving an highly heterogeneous picture of intermediate metabolism.
This is consistent with the time-hierarchy hypothesis, by which typical metabolic timescales should be highly heterogeneous in order to suppress instabilities\cite{park1974hierarchical, reich1975time, heinrich1978metabolic}.
As we mentioned in the introduction we can go a step beyond  and calculate forward and backward contribution of reaction fluxes by using the fluctuation theorem from the knowledge of the reactions' free energies. These can be estimated by performing an energy balance analysis of the network. Reactions free energies should be consistent with the reaction directions:
 $ \textrm{sign}(v_i) \Delta G_i \leq 0$ and if we  decompose them in terms of metabolites chemical potentials $\Delta G_i =\sum_\mu S_{i \mu} g_\mu$, we have for given reaction directions the feasible space of chemical potentials:
\begin{eqnarray}
\xi_{i \mu} = \textrm{sign}(v_i) S_{i \mu}  \\
\sum_\mu \xi_{i \mu} g_\mu \nonumber \leq 0 
\end{eqnarray}
from which free energies can be inferred with relaxational methods starting from a reliable experimental prior.

Once a free energy vector has been retrieved,
the backward and forward contributions to fluxes can be calculated from the fluctuation theorem:
\begin{eqnarray}
v_i^+ = \frac{v_i}{1-e^{\Delta G_i/RT}} \nonumber \\
v_i^- =  \frac{v_i}{e^{-\Delta G_i/RT}-1} \nonumber \\
\end{eqnarray}
The total fluxes are reported  in  fig 1 (bottom)  in log scale comparing them with net values. 
\begin{figure}[h!!!!!]\label{fig2}
\begin{center}
\includegraphics*[width=0.7\textwidth,angle=0]{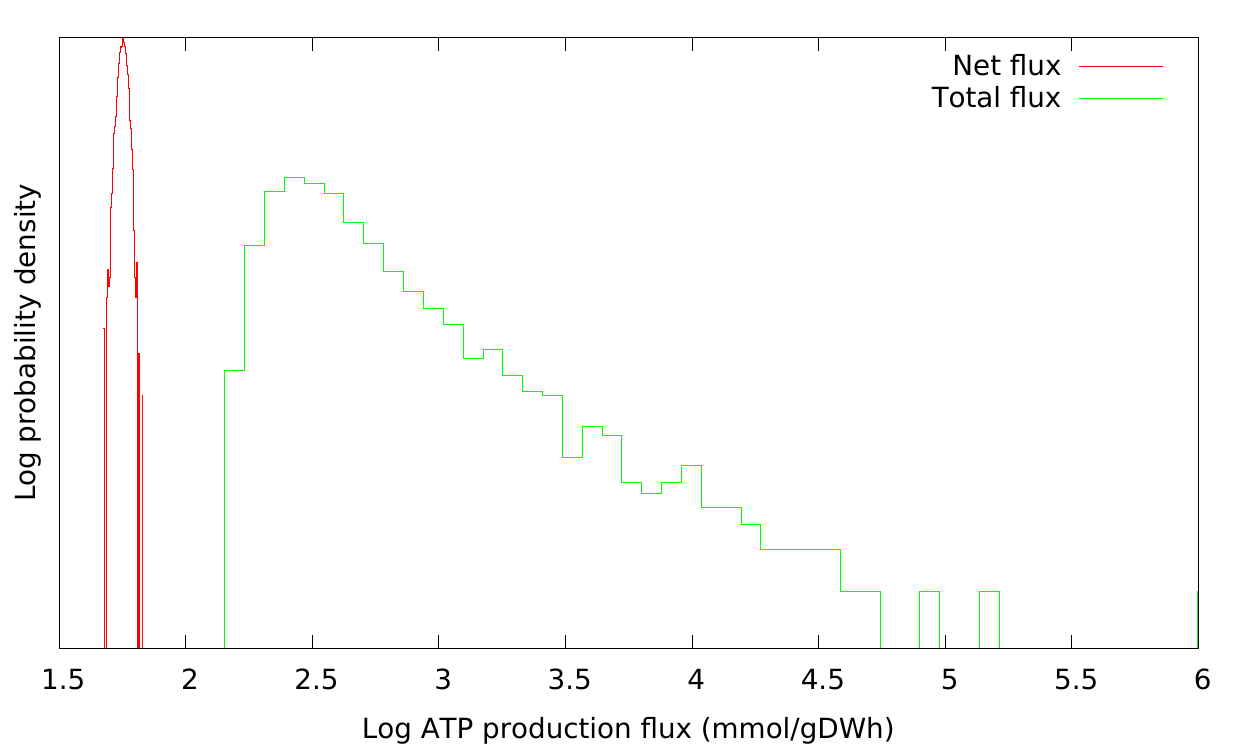}
\includegraphics*[width=0.7\textwidth,angle=0]{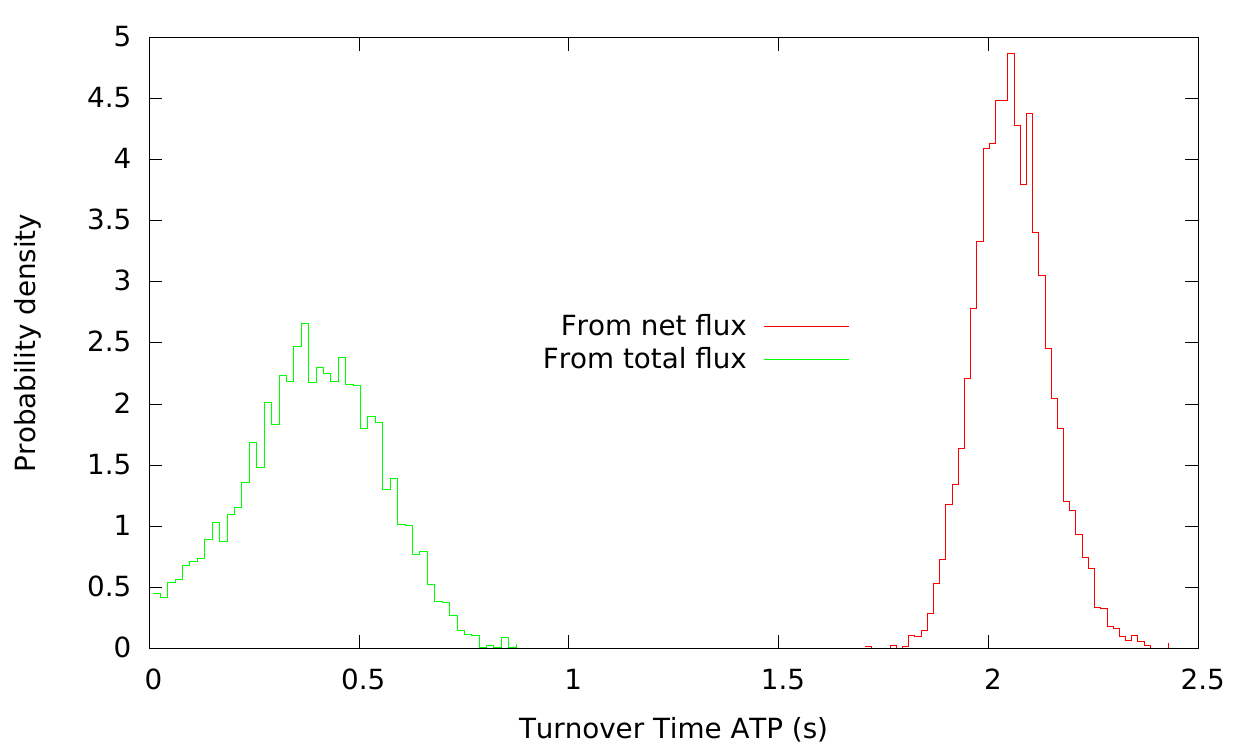}
\caption{Top: net and total ATP production flux  distributions in the metabolic network model of E Coli iJR904 in a glucose-limited minimal medium. Bottom: inferred ATP turnover time distributions.}
\end{center}
\end{figure}

We can see that for many metabolites the corrected production flux can be one order of magnitude higher when it is processed by enzymes working nearby thermal equilibrium, that would imply faster relaxation times of the system, for example in its response to perturbations.
In figure 2 (top) we plot for instance the distribution of ATP production fluxes  in log scale: we pass from a distribution centered around the mean with value $56$mmol/gDWh to an heterogeneous distribution  with a fat tail and peaked around $350$mmol/gDWh. 
This correction lead to different qualitative estimate of the relative turnover time: in fig 2 (bottom) the distribution of the turnover time of ATP inferred from net and total fluxes assuming a concentration of $9.6$mM\cite{bennett2009absolute} is reported: we pass from a mean of $2.0$s (consistent with previous estimates reported in databases\cite{milo2010bionumbers}) to $0.4$s upon using the correct  value from total fluxes.

In fig 3 we plot in log scale turnover times from net and total fluxes for several metabolites for which we have used the measures of concentration of\cite{bennett2009absolute} 
\begin{figure}[h!!!!!]\label{fig5}
\begin{center}
\includegraphics*[width=1\textwidth,angle=0]{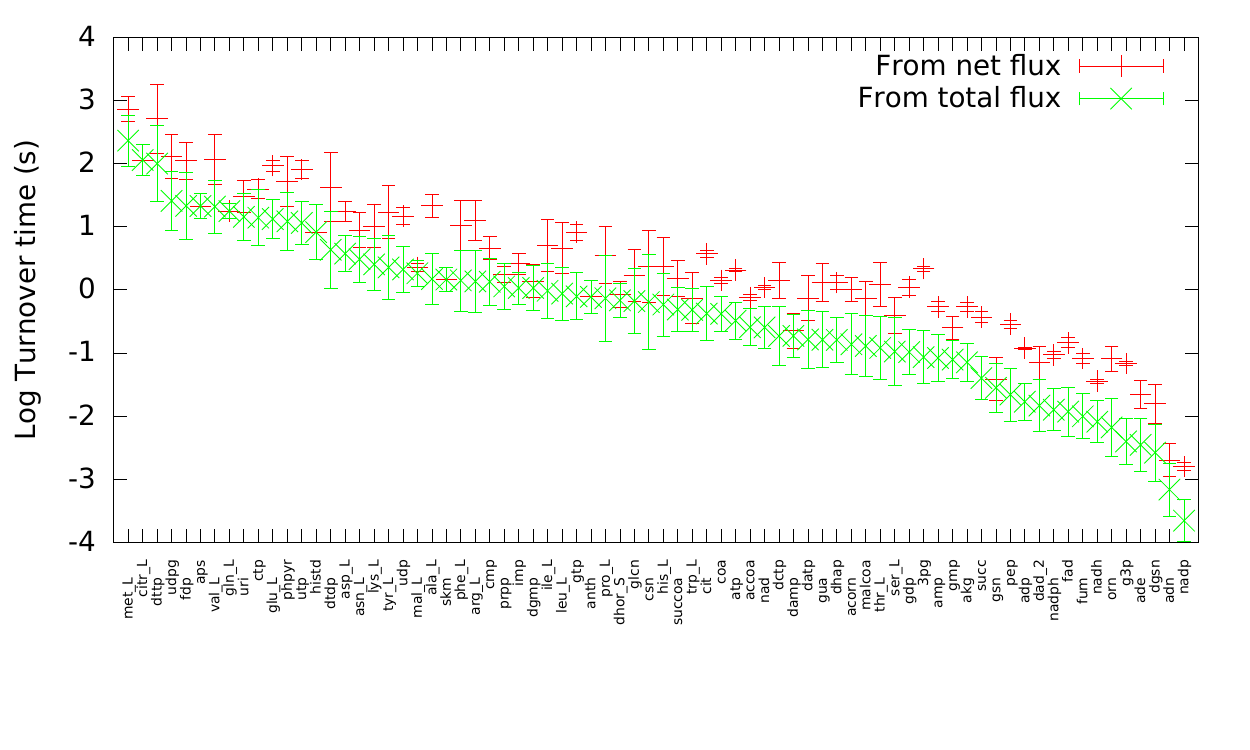}
\caption{Turnover time of several metabolites inferred from net and total fluxes.}
\end{center}
\end{figure}
\begin{table}[h!!!]
\begin{tabular}{  | c  | c  |  c | }  
\hline 
Metabolite  & Tunover time from net flux (s) & Turnover time from total flux (s)  \\ 
\hline 
Glutammate  &  $90 \pm 20$  & $16 \pm 8$   \\
3-Phosphoglycerate  &  $2.0\pm0.2$  & $0.12\pm0.6$   \\
ATP  &  $2.0\pm0.1$  &  $0.4\pm0.2$  \\
ADP  &   $0.120\pm0.005$ & $0.02\pm0.01$   \\
AMP  &  $0.5\pm0.1$  & $0.11\pm0.06$  \\
NAD  &  $1.1\pm0.1$  & $0.3\pm0.1$  \\
NADH  &  $3.5\pm0.2 \cdot 10^{-2}$  & $1.0\pm0.4 \cdot 10^{-2}$   \\
NADP  &  $1.6\pm0.2 \cdot 10^{-3}$   & $2\pm1 \cdot 10^{-4}$    \\
NADPH  & $9\pm1 \cdot 10^{-2}$    &  $2\pm1 \cdot 10^{-2}$   \\
\hline
\end{tabular}
\caption{Turnover time estimates from net and total fluxes}
\end{table}
We present in table 1 for comparison the turnover times estimate from net and total fluxes of some key metabolites.

\section{Materials and methods}

\subsection{Data}
The network  employed is E Coli iJR904\cite{reed2003expanded}.
This network consists of $N=1075$ reactions among $M=761$ metabolites.  Upon considering a glucose limited minimal medium we are left, after removing blocked reactions and leaf metabolites with $N=667$ reactions among $M=450$ metabolites.
We set a maximal glucose uptake of $u_g=-6$mmol/gDWh and fix a minimal ATP maintenance of $v_{ATP}=7.6$mmol/gDWh.
The resulting polytope has dimension $D=233$.
In regard to the data for the energy balance analysis  we recur  to the chemical potential estimates coming from a genome scale application of the group contribution methods from\cite{Jankowski:2008bh,Henry:2006mi}.
In order to calculate turnover times we have recurred to the measure of absolute concentrations reported in\cite{bennett2009absolute}.
\subsection{Computational methods}
\subsubsection{Uniform sampling of steady states}
An uniform sampling of a convex polytope in high dimension is usually performed with  Markov chain Monte Carlo methods, since an exact enumeration of the vertex would be infeasible due to their exponential number and static rejection methods\cite{Price:2004p4298} would suffer as well from high-dimensionality issues.
We have recurred to well known  hit and run markov chain\cite{Turcin:1971} that has been investigated in much detail in mathematical literature, showing nice convergence properties: it is guaranteed by detail balance to converge to the uniform distribution with a  mixing time that scales as\cite{Lovasz:1999p4121} 
\begin{equation}
\tau = O\left(D^2 (\frac{R}{r})^2\right)
\end{equation}
where $r$ and $R$ are the radius of respectively the biggest (smallest) inscribed (inscribing) sphere.
The factor $R/r$ ('sandwitching ratio') can lead to ill-conditioning issues, since the timescales of metabolic fluxes are typically very heterogeneous.  Such factor can be reduced to a polynomial of the  space dimension with a polynomial time algorithm that finds a rounding ellipsoid\cite{lovasz1987algorithmic}.
We have rounded the polytope with an ellipsoid founded with the flux variability analysis\cite{Mahadevan:2003pi}, that has been shown to reduce the sandwitching ratio to values that allow an efficient sampling\cite{uniformell}  
\subsubsection{Energy balance analysis}
Finding a free energy vector consistent with a given flux configuration amounts at solving a system of linear inequalities for which relaxational algorithms can be employed\cite{minover0,Motzkin:1954p4304} starting from the experimental prior that we have reported in the subsec. Data. However we point out that duality arguments\cite{deMartino12, muller2012thermodynamic} lead to the infeasiblity of such system if closed reactions loops are present. Such closed reaction loops  has been exhaustively enumerated in this network  and were corrected in a minimal way\cite{de2013counting}.

\section{Conclusions}
In this work we have revisited the notion of metabolic turnover in the light of recent results in out-of-equilibrium thermodynamics.
The net flux of  a reaction working nearby thermodynamic equilibrium result from a contribution of backward and forward fluxes, whose value can be inferred from the fluctuation theorem upon knowledge of the free energy.
The higher resulting total fluxes lead to effectively faster relaxation times in metabolic systems, a notion that can be captured at an approximated level by the computation of metabolites turnover times. We have performed an uniform sampling of the steady states of an E Coli genome scale constraint based metabolic network model, we have peformed an energy balance analysis of the network and we have estimated total production fluxes from the fluctuation theorem.
We have shown that metabolites turnover times estimated in this way can be as far as one order of magnitude lower than the ones inferred from net fluxes. 
Such reduction of turnover times could lead in principle to values that are below typical metabolites diffusion times
and thus it could affect the well-mixing hypothesis, that is at the core of our approach and constraint-based modeling in general. An approximate estimate\cite{milo2010bionumbers} in which we consider a
diffusion constant of  $D\simeq 200$ $\mu m^2$/s, and  that the diameter of E coli is $d\simeq 1 \mu m$ , lead to an order of magnitude estimate $t_{d}\simeq \frac{d^2}{6 D} \simeq 10^{-3}$s. The turnover time of all the metabolites we have calculated is above this threshold apart from NADP and adenosine. Such analysis could be performed in principle for any metabolic network  upon knowledge of the free energies landscape. On the other hand a more rigorous estimate of  the true relaxation times of a metabolic system would requires genome-scale insights on enzymatic kinetic laws, including allosteric regulations, an aspect that would require further investigations.

\section*{Acknowledgments}
The author thanks A. De Martino for interesting discussions. This work is supported by the DREAM Seed Project of the Italian Institute of Technology (IIT). The IIT Platform Computation is gratefully acknowledged.
\section*{References}    
\bibliographystyle{unsrt}
\bibliography{turnbib}




\end{document}